\documentstyle[prl,aps]{revtex}
\large
\begin{document}
\twocolumn
\draft
\title{Wentzel--Bardeen singularity in coupled\\
Luttinger liquids: transport properties}
\author{Thierry Martin$^*$}
\address{Theoretical Division, CNLS, Los Alamos National Laboratory,
Los Alamos, NM 87545}
\date{\today}
\maketitle
\begin{abstract}
The recent progress on 1 D interacting
electrons systems and their applications to study the transport properties of
quasi one dimensional wires is reviewed. We focus on
strongly correlated electrons
coupled to low energy acoustic phonons in one dimension.
The exponents of various response functions are calculated, and their
striking sensitivity to the Wentzel--Bardeen singularity
is discussed.
For the Hubbard model coupled to phonons the equivalent of a phase diagram is
established.
By increasing the filling factor towards half filling the
WB singularity is approached. This in turn
suppresses
antiferromagnetic fluctuations and drives the system
towards the superconducting regime, via a new intermediate (metallic)
phase.
The implications of this phenomenon on the transport properties
of an ideal wire as well as the properties of a wire with weak or strong
scattering
are analyzed in a perturbative renormalization group calculation.
This allows to recover the three regimes predicted from the
divergence criteria of the response functions.
\end{abstract}
\smallskip
\bigskip
\narrowtext
\section{Introduction}

The developments in the field of Mesoscopic physics have generated a lot of
excitement for more than a decade. One of the reasons for this is
connected to the recent advances in fabrication techniques, which have allowed:
a) the discovery of new phenomenon associated with quantum interference effects
in small structures b) the verification of a plethora of
theoretical predictions on the same systems.
This relative balance between theory and experiment is very desirable
in any field of physics.
Apart from a few exceptions \cite{Rice,Altschuler,Ambegaokar}, it seems fair to
say that until recently, the theoretical effort in Mesoscopic physics has
dealt mostly with non interacting electrons: phenomenon associated with sample
to sample averaging, interference effects in small disordered wires
and rings,
transport properties from a scattering matrix approach,
chaotic phenomena in Mesoscopic conductors, to cite a few.
The Coulomb blockade certainly deals with electron interactions, however,
for large quantum dots,
it can be argued in terms of classical capacitance effects \cite{Coulomb},
as opposed to a first principles treatment of interactions between electrons.
There are several ``kinds'' of quasi one dimensional (1 D) systems:
a) systems obtained by lateral confinement of a 2 D electron gas
into a narrow channel, b) micrometer metallic wires with
a small aspect ratio, c) systems which have an intrinsic 1 D
character like edge states in the integer (IQHE) and fractional (FQHE)
quantum Hall effect, and of course, d) molecular chains and
bundles of all sorts which have been studied for a long time.
In the latter, rigorous methods from the 1970's \cite{Emery,Solyom}
have long been applied to predict ``phase'' transitions
from charge/spin density waves to singlet and triplet superconducting
ground states as various coupling parameters are varied.
By ``phase'', it is implied which type of fluctuations
dominate, as we know that long range order is
strictly not possible in 1 D. For ``artificially tailored'' systems
in categories a) and b), disorder remains a technological challenge
to this day. On the contrary, edge states in the quantum Hall regimes
( c) ) have the advantage that the effect of impurities can be minimized
because of the absence of backscattering.

Here we study a 1 D interacting electron system coupled to a
low energy bosonic field. This system exhibits the interplay
beween the Coulomb repulsion and the attractive, retarded interaction
mediated by the phonons. By varying the electron density alone,
it is possible to reach a strong electron--phonon regime with
dramatic consequences on the thermodynamic properties.
Phase transitions in this system
will be considered, and some emphasis will be put on the
transport properties of ideal
and disordered wires in the presence of coupling to a bosonic field.
In practice, this boson field will be chosen to represent the phonons
of the underlying lattice, but this is by no means a serious constraint.
The bosonic field can just as well represent the charge field of
another 1 D electron branch.

In 1D, the Fermi liquid ground state becomes unstable,
as the interactions
drive the systems into a new phase called the Luttinger liquid
phase \cite{Mattis,Luther,Haldane}. The jump of the momentum
distribution function at the Fermi surface which is characteristic of
Fermi liquids vanishes, leaving only
a branch cut singularity at this point.
The collective excitations of the electron near the
Fermi level can adequately be described by a set of
harmonic oscillators.
At low temperatures, the single electron
spectrum can be approximated by a linear spectrum, and considerable
simplifications follow. The Fermion field operators are decomposed into right
and left moving components, which in turn can be expressed
as exponentials of bosonic fields chosen to reproduce the anti--commutation
relations. One is then left with scalar, bosonic fields, which
are particularly suited for a functional integral
representation of the partition function. Moreover, a renormalization
group analysis can be coupled to this approach in order to determine which
(and when) non linear operators are relevant. In the end (for specific range
of electron density, coupling parameters) one has to deal essentially
with an action which is
quadratic in the bosonic fields, with parameters which are renormalized
by the (irrelevant) operators.

The transport properties for a wire with ideal transmission yield a
conductance which is no longer quantized in units of $e^2/h$, the
non interacting result. Rather, the Luttinger liquid model predicts
\cite{Kane,Furusaki}
$G=2e^2K_\rho/h$ {\it below} the free electron value, as $K_\rho<1$
for electrons with repulsive interactions.
In the presence of a
single barrier the conductance vanishes as $T^{2/K_\rho-2}$. This
applies for temperatures above the cutoff
$T_c\equiv \hbar v_F/L k_B$, where $L $ is the length of
the 1 D region. Below this cutoff, effects associated with
the massive leads connected to the wire provoke a saturation of
the conductance to a finite value.
Despite this dramatic vanishing of the temperature,
perfect transmission in two--barrier
resonant tunneling structures can be achieved for a wide
range of electrons
couplings by tuning the
gate voltage on the
central region, because of constructive interference
effects associated with the two barriers \cite{Kane,Furusaki}.
For resonant tunelling in the FQHE regime \cite{Moon}
the resonance line shape is {\it universal}
and deviates strongly from the Fermi liquid behavior. For free
electrons, the line shape is supposedly Lorentzian and does
not depend on the temperature. On the contrary, for
a Luttinger liquid the peaks in conductance are delta functions at $T=0$
and widen at finite temperature like $T^{1-\nu}$, with $\nu=1/m$
($m$ integer) the filling factor. This remarkable result
was recently tested experimentally \cite{Webb} in a FQHE experiment
with lateral confinement. Non--equilibrium current or voltage noise
across a weak point contact in the FQHE has been shown
(theoretically) to provide a direct measurement
of the fractional charge: for a constriction with good transmission
properties, the noise originates from the infrequent tunneling of Laughlin
quasiparticles \cite{Kane noise}.
The Kondo effect in 1 D interacting electron systems has been studied by
several authors \cite{Kondo}. Persistent currents
for interacting electrons in the ballistic regime have been
predicted \cite{Loss,LossM}.

1 D electron systems coupled to small-momentum phonons
have been considered on several occasions. 40 years ago,
Wentzel \cite{Wentzel} and Bardeen\cite{Bardeen} used a similar model
as a tentative candidate to explain the theory
of conventional superconductivity. For a critical
electron--phonon coupling
constant, the system becomes {\it unstable}, and the specific heat
diverges as one approaches this critical point \cite{Engelsberg}:
the retarded interaction
mediated by the phonons induces a collapse of the system.
Below, we refer to this singular point as the
Wentzel--Bardeen (WB) singularity. Nevertheless,
this mechanism was not considered too seriously,
perhaps because of the
relatively large values of the electron--phonon coupling constant
needed to reach the singularity? Also, it is rather difficult
in an experiment to
tune this coupling constant to a desired value.
An exactly solvable model similar to that of Ref. \cite{Engelsberg} was
studied recently \cite{Voit0}, with the conclusion that
these small-momentum phonons can be ignored in typical metals since
their effect is of the order $c^2/v_F^2\ll 1$ ($v_F$ is the Fermi
velocity and $c$  the sound velocity).
Traditionally \cite{Apostol},
thermodynamic quantities have been expanded
in terms of this ratio.
Consequently, the interest \cite{Voit1,Zimanyi,Voit2}
shifted towards larger-momentum ($2k_F$) processes,
where an electron is scattered by a phonon across the Fermi surface.

The point we wish to make is that when the electron system
includes Umklapp and backscattering processes,
the small-momentum phonons
{\it do} play an
important role nevertheless.
It is no longer the ratio $c/v_F$ which
characterizes the corrections associated with the phonons.
Rather, the velocity $u_\rho$
associated with particle--hole
excitations close to
the Fermi level replaces the Fermi velocity $v_F$.
The Fermi velocity still determines the energy scales of the electron
Hamiltonian. It is only the fluctuations near the Fermi level which
are described by $u_\rho$.
For the Hubbard model in particular, $u_\rho$ vanishes
at half filling due to the presence of the insulating phase.
Consequently, the coupling to small-momentum transfer modes
(such as acoustic phonons)
is a {\it non-perturbative} effect in the ratio $c/u_\rho$ as
one reaches the WB singularity.
For the Hubbard model, the WB singularity is accessed
for arbitrary electron--phonon
coupling constant by mearly increasing the filling factor
towards half filling.
Near this singularity, CDW and SDW fluctuations are suppressed
totally,
and the system is pushed into a metallic  and finally
superconducting phase by slightly increasing the filling factor.
The presence of an intermediate phase between the SDW/CDW and
superconducting regimes constitutes a novelty.
\section{model}
\label{model section}

Our starting point  is the Hamiltonian $H=H_e+H_{p h}+H_{e-p h}$, where
\begin{equation}
H_{p h}={1\over 2}\int d x~[\zeta^{-1}\Pi_d^2+ \zeta c^2 (\partial_x
d)^2]
\label{phonon}\end{equation}
describes the free phonons, with $d$ the
displacement field, $\Pi_d$ its canonical
conjugate and $\zeta$ the mass density.
The electron--phonon coupling is
\cite{Fetter}:
\begin{equation}
H_{e l-p h}=g\sqrt{\pi\over 2}\sum_s\int d
x~\psi_s^{\dagger}\psi_s \partial_x d~,
\label{coupling}\end{equation}
where $g$ is the coupling constant, and $\psi_s=
e^{i k_Fx}\psi_{s+}+e^{-i k_Fx}\psi_{s-}$
is the electron field operator. In terms of boson fields,
$\psi_{s\pm}({\bf x})\propto\exp\bigl[i\sqrt{\pi}\bigl(\pm \varphi_s({\bf
x})-\int^x\Pi_s(x^\prime,\tau)dx^\prime)\bigr)\bigr]$ (here ${\bf x}=(x,\tau$))
representing right and left moving electrons.
The slow spatial variation of the electron density in Eq. (\ref{coupling})
reduces to $\sum_s\psi_s^\dagger
\psi_s=\sqrt{2/\pi}\partial_x\varphi_\rho$, with
$\varphi_{\rho, \sigma}=(\varphi_\uparrow\pm\varphi_\downarrow)/\sqrt{2}$
the charge and spin fields (similarly for the canonical
conjugate fields $\Pi_{\rho, \sigma}$). We neglect the
fast oscillating terms in the density which can be compensated
by the $2k_F$ phonons. Therefore,  phonon mediated backscattering
which transfers an electron from one side of the Fermi ``surface''
to the other is not included here, but (non retarded) backscattering
from electron electron collisions is otherwise included.
Later on in this paper, we will show that the present model can also be applied
to electron wave guides, where different modes associated with lateral
confinement interact with each other. The problems associated with the $2k_F$
phonons are not present there.
Following Ref. \cite{Schulz}, the electronic Hamiltonian is
$H_{e l}=H_\rho+H_\sigma$, with
\begin{equation}
H_\nu={1\over 2}\int d x~\Biggl[u_\nu K_\nu\Pi_\nu^2
+{u_\nu \over K_\nu}(\partial_x\varphi_\nu)^2\Biggr]~,
\label{nu Hamiltonian}\end{equation}
with $\nu=\rho, \sigma$. In principle, non--linear terms should be added
to $H_{e l}$. However, below half filling, these terms are irrelevant
and can be accounted for by renormalizing the parameters
$K_\rho$, $K_\sigma$, and the charge
(spin) velocities $u_\rho$ ($u_\sigma$).
The phonons couple only to the charge degrees of freedom,
and this property is preserved in the propagators calculated
below. For free electrons,
$u_\rho=u_\sigma=v_F$ and $K_\rho=K_\sigma=1$.

\section{Correlation functions}
\label{correlation section}

To calculate thermodynamic quantities, we use
a functional integral representation of the partition function.
As we will be calculating  Green functions which involve
fermion operators only, the phonon degrees of freedom can be
integrated out right away. The resulting effective action for
the charge degrees of freedom, reads in  Fourier representation:
$S_\rho=(2\beta L)^{-1}\sum_{\bf k}D_\rho({\bf k})|\varphi_\rho({\bf
k})|^2$ (${\bf k}=(k,\omega)$), with the inverse propagator:
\begin{equation}
D_\rho({\bf k})={1\over K_\rho u_\rho}\biggl(\omega^2+u_\rho^2 k^2
-{b^2 k^4\over \omega^2+c^2k^2}\biggr)~,
\label{propagator}\end{equation}
where $b=g\sqrt{K_\rho u_\rho /\zeta}$. The first two terms
of Eq. (\ref{propagator})
represent the contribution of the free charge field, and the
retarded, attractive coupling associated with the phonons appears in
the third term.
At the WB point the charge density propagator,
$k^2/D_\rho({\bf k})$, becomes proportional
to $\omega^{-2}$ and signals an instability
towards long wave--length density fluctuations.

\subsection{Green function}
\label{Green subsection}

We first consider the single--particle Green function:
\begin{equation}
G_s(x,\tau)=-<T_\tau\psi_s(x,\tau)\psi_s^\dagger(0,0)>~,
\label{Greens 0}\end{equation}
where $T_\tau$ is the imaginary time ordering operator.
Using the decomposition of the Fermi operators into right and
left moving components, and after normal
ordering, the calculation of the Green function
is reduced to an evaluation of Gaussian integrals.
Analytic expressions for the Greens function can be obtained for this
particular model, and are found in Refs.
\cite{Loss Martin,Martin Loss}.
The coupling between the charge degrees of freedom and the phonons
induces an hybridization of the two excitations with characteristic
velocities:
$v_\pm^2=[u_\rho^2+c^2\mp\sqrt{
(u_\rho^2-c^2)^2+4b^2}]/2$
At large distances, the Green function decays as a power law,
\begin{equation}
G_s(x,0)\propto |x|^{-1-\delta}~,
\end{equation}
with
$\delta=K_\sigma/4
+1/(4 K_\sigma)-1
+B/(4K_\rho)+A K_\rho/4$.
The electron--phonon parameters are defined by
$A= u_\rho(1+c^2/v_+v_-)/(v_++v_-) \ge 1$, and
$B= u_\rho(1+v_+v_-/u_\rho^2)/(v_++v_-)\le 1$.
These parameters play a crucial role in the discussion
of the ordering fluctuations below. In the limit $g\rightarrow 0$,
$A=B=1$. Let us now consider the case $g\neq 0$, where
$A>1$ and $B<1$\cite{Martin Loss}:
As $u_\rho/K_\rho$
approaches the critical value
$u_\rho^*/K_\rho^*=g^2/(\zeta c^2)$ from above,
$v_+^2$ tends to zero, and $v_-^2$ to ${u_\rho^*}^2+c^2$.
As a result, the exponent $A$ diverges and  $B$ decreases
to the finite value $B^*=u_\rho^*/\sqrt{{u_\rho^*}^2+c^2}<1$.
For $u_\rho/K_\rho< u_\rho^*/K_\rho^*$ the velocity $v_+$ becomes
complex and the model becomes unphysical. Thus we must require
that $u_\rho/K_\rho\ge u_\rho^*/K_\rho^*$, or equivalently that
$b/(cu_\rho)\leq 1$, the equality sign defines the WB
singularity \cite{footnote1}.
Because $\delta>0$, if $g\neq 0$ \cite{Martin Loss}, the Green
function decays faster than $1/x$, the free
electron result. From the large distance behavior, we find the
momentum distribution function near $k_F$,
$N(k)\simeq N(k_F)-\kappa {\rm sgn}(k-k_F)|k-k_F|^{\delta}$,
with $\kappa$ some constant of order one, and
$N(k_F)=\Gamma(1/2+\delta/2)/[2\sqrt{\pi}\Gamma(1+\delta/2)]$.
The jump at the Fermi surface disappears.
The presence of phonons induces this Luttinger liquid
behavior even without  electron-electron interaction
($K_\rho=K_\sigma=1$).

We plot the quantity $\delta$ for the Hubbard model
(see Sec. \ref{Hubbard section}) coupled to phonons
in Fig. \ref{fig1}a
and \ref{fig1}b.
We first plot $\delta$ for a quarter filled band as a
function of the on site repulsion $U$ in Fig. \ref{fig1}a,
for several values of
the phonon coupling parameter. For small phonon coupling, $\delta$
increases monotonically with increasing $U$. As the phonon coupling
is further increased, $\delta$ acquires a minimum.
This is an early signature of the
competition between these two couplings, and will be more apparent in the
discussion of the phase diagram.

Next, we choose $U=2$ and vary the filling factor from zero to half
filling in Fig. \ref{fig1}b. $\delta$ increases dramatically when these
two extremes are reached, as the correlation effects between electrons
dominate the physics in both cases. In particular,
$\delta\rightarrow\infty$ at the WB singularity.

\subsection{Ordering fluctuations}
\label{2 correlation}

The definitions for the ordering correlation functions $N({\bf x})$,
$\chi({\bf x})$, $\Delta_s({\bf x})$ and $\Delta_t({\bf x})$
which describe CDW, SDW, singlet (SS) and triplet (TS)
superconducting fluctuations are given in Refs.
\cite{Emery,Solyom,Martin Loss}.
At large times/distances, these response functions behave like
power laws.
The signature for a particular ordering fluctuation to be present is
given by the divergence of the Fourier transform of the corresponding
response function at low frequency and  small momentum
relative to $q=2k_F$ ($q=0$) for $N$ and $\chi$ ($\Delta_{s, t}$).
We thus obtain the following criteria \cite{Loss Martin,Martin Loss}:
\begin{mathletters}
\begin{eqnarray}K_\rho A+K_\sigma&\leq&
2~~~~~~~{\rm (CDW)~~~~}\label{criterion
C D W} \\
K_\rho A+1/K_\sigma&\leq&2~~~~~~~{\rm
(SDW)~~~~}\label{criterion S D W}\\
B/K_\rho+K_\sigma&\leq&2~~~~~~~{\rm (SS)}
\label{criterion singlet}\\
B/K_\rho+1/K_\sigma&\leq&2~.~~~~~~{\rm
(TS)}\label{criterion triplet}
\end{eqnarray}\end{mathletters}
Since $A>1$ and $B<1$ for $g\neq 0$, we see that the Cooper instability
is always present for non--interacting
electrons coupled to phonons.

\subsection{Hubbard model}
\label{Hubbard section}

We specialize to a Hubbard model for the correlated electrons
with on site repulsion $U$ and hopping term $t$.
In this case $K_\sigma=1$ because of SU(2) symmetry, and as a result
the CDW and the SDW (the SS and the TS) response functions have
apparently the same exponents (except at half filling).
However, logarithmic corrections originating from marginally
irrelevant operators in the spin channel favor SDW (TS) over
(CDW) SS fluctuations \cite{Logarithm}.
The remaining parameters $u_\rho$ and $K_\rho$ of the Luttinger
liquid Hamiltonian are obtained for arbitrary on--site repulsion
$U$ and filling factor $n$ by solving two integral equations
\cite{Schulz,Martin Loss} describing the ground state properties \cite{Lieb}
and the spectrum of charge excitations \cite{Coll}.

Our results are plotted in Fig. \ref{fig2} a) and b) where we have
chosen $c=a_0t/\hbar$ ($a_0$ is the lattice constant):
for fixed $U/t$ (Fig. \ref{fig2} a), we determine
which fluctuations dominate as a function of $n$ and an effective
electron--phonon parameter $b/(u_\rho c)$: for
convenience we  consider in these plots $n$ ($U$) and
$b/(u_\rho c)$ as independent parameters.
For small $b/(u_\rho c)$, SDW (i.e. antiferromagnetic) ordering fluctuations
dominate for
arbitrary filling factor. As this parameter is increased,
we reach (away from half filling) a region for which no
response function diverges: we refer to this ``phase''
as the metallic region. At low $U$ and
$b/(u_\rho c)\ll 1$ analytical results confirm the existence
of this intermediate phase \cite{Martin Loss}.
By further increasing $b/(u_\rho c)$, the region where superconducting
fluctuations dominate is reached. On the other hand, near
half filling,
the correlation effects suppress the
superconducting phase.
For larger values of $U$, the superconducting
region shrinks towards the region where $b/(u_\rho c)=1$,
and the SDW region grows as expected \cite{Martin Loss}.

Next, we choose quarter filling ($n=1/2$), and plot in Fig. \ref{fig2} b) the
phase diagram as a function of $U$ and the phonon coupling.
At low $U$, $K_\rho\approx 1$, and the system is superconducting
because of the Cooper instability. As $U$ is further
increased,  one crosses the metallic and the SDW (CDW) phase,
because the phonon--mediated attractive
interaction is overcome by the instantaneous repulsion between electrons.

\subsection{WB singularity}
\label{Wentzel subsection}

In both Figs. \ref{fig2} a) and b), the upper line
$b/(u_\rho c)=1$ corresponds to the WB singularity. We now
discuss the relevance of this singularity for the Hubbard model.
Plotting $u_\rho/K_\rho$ for several values of $U$ as a function
of the filling factor (Fig. \ref{fig3}), we notice
that $u_\rho /K_\rho$
vanishes as one approaches half filling.
(for small $U$, the
curve develops a peak located close to $n=1$, and the fall to zero
is all the more drastic).
Thus, the WB singularity
at $u_\rho^*/K_\rho^*$ can be reached
{\it for arbitrary values of the coupling constant}
$g$ ($\neq 0$). Near half filling, as one approaches
$u_\rho^*/K_\rho^*$ from lower
filling factors, the divergence in $A$
triggers a dramatic suppression of the SDW (CDW) fluctuations,
and the system is driven into the metallic phase \cite{footnote}.
Moreover, since $B\rightarrow B^*$ as $u_\rho\rightarrow u_\rho^*$, and
$B^*\propto g^2$, the SS/TS condition $B\leq K_\rho$ can be met near
half filling for sufficiently small $g\neq 0$
(since $1/2\leq K_\rho\leq 1$).
Hence, the system can finally be driven into the superconducting phase
by approaching half filling.
In summary, the conjunction of the coupling to small momentum
acoustic modes with the reduction of the charge velocity near
half filling provides a very efficient mechanism to suppress SDW (CDW)
order, and to drive the system into a metallic and
finally superconducting phase. This mechanism
need not be limited to the
1D Hubbard model, but could occur in other situations
where strong correlations play an important role.

We note that this change of phases as a function
of filling factor is reminiscent
of the behavior of high $T_c$ materials under doping.
Since it is the low--momentum and not the $2k_F$
phonons which are important here
and since strong correlations are also present in 2D,
it might not be unreasonable to hypothesize
that the mechanism discussed here is also realized in higher dimensions.

\section{Application to electron wave guides}
\label{analogy}

We now illustrate our results with an
application to
Mesoscopic quasi--1D wires.
It is well known that the transverse states of the electron
wave function in a 2 D electron gas with a point contact
are at the origin of the conductance quantization in point contact
experiments \cite{Van Wees}.
In a 1 D wire, one can also adjust how many of these
transverse states are populated by tuning the voltage on an
electrostatic gate.

\subsection{Interaction between modes}

Consider the Coulomb interaction between two different branches:
if $d$ is the distance from the 2D electron gas to the metallic gate, this
interaction is screened for distances $x>d$.
The interaction potential between mode ``i'' and mode ``j''
has the form:
\begin{equation}
V_{i j}=\sum_{s, s^\prime}\int dx\int dx^\prime
\rho_{is}(x)V(x-x^\prime)\rho_{js^\prime}(x^\prime)~,
\label{potential i j}
\end{equation}
with $\rho_{is}$ the density operator for electrons of mode
$i$ and spin $s$, and $V(x-x^\prime)$ the screened Coulomb
potential. As we are concerned with low temperature, long wave length
properties, we can replace this interaction by a delta function potential
$V_0\delta(x-x^\prime)$ with \cite{Matveev}.
\begin{equation}
V_0\simeq 2 {e^2\over \epsilon}\ln(k_Fd)~,
\label{delta coefficient}
\end{equation}
where $\epsilon$ is the dielectric constant of the medium between
the 2D gas and the gate.
The global Hamiltonian  describing the set of coupled Luttinger liquids
is therefore:
\begin{equation}
H=\sum_{i}(H_{i\rho}+H_{i\sigma})+g \sum_{i j}
\int dx (\partial_x\varphi_{i\rho})(\partial_x\varphi_{j\rho})~.
\label{coupled Hamiltonian}
\end{equation}
The first term corresponds to the set of uncoupled charge and spin fields
associated with each modes, and the coupling term has striking
similarities with the electron--phonon case. The phonon
field we had previously is simply replaced by the charge field
associated with another mode, but the spin degrees of freedom are uncoupled
(as long as there are no magnetic impurities).
A similar model was used in Ref. \cite{Matveev}
in the discussion of the transport properties of a multimode wire
through a point contact, and is equivalent
to the multicomponent
Tomonaga Luttinger model \cite{Penc}. Also, the present interaction
does not allow a transfer from one mode to the other, but rather the
direct interaction as well as the exchange interaction between
these modes. We come back to this point below.

Note that the fast spatial variation ($2k_F$) of the electron density
can be safely neglected here. There is no compensation between two
rapid density oscillations in analogy with $2k_F$ phonons,
because each mode has a fixed Fermi velocity specified by the lateral
confinement and the electron density.

Next, we choose the simplest situation:
an electron density such that only two modes, the transverse
ground state and the first excited state, are present. This system
is now an exact analog of the electron--phonon situation, because the
spin degrees of freedom of each modes are uncoupled.
Consequently, by increasing the coupling between the two modes, we
can expect to induce superconducting fluctuations in one or both branches.
The coupling can be varied in several ways: by considering several samples
with different separation between the 2D gas and the metallic gates,
by increasing the electron density in the fundamental mode to the
analog of half filling, or by lowering the density such that
the Fermi velocity of the first excited state is adequately close to zero.
The latter two methods exploit the
WB singularity.

The Green function exponent associated with electrons from chain $i$
in then given by
$1+\delta_i=K_{\sigma i}/4+1/4K_{\sigma i}+
B_i/4K_{\rho i}+K_{\rho i}A_{i j}/4$. In analogy
with the phonon case, we have defined
$A_{i j}=u_{\rho i}\biggl(1+u_{\rho j}^2/
(v_+v_-)\biggr)/(v_++v_-)$, $B_i=u_{\rho i}
\biggl(1+v_+v_-/u_{\rho i}^2\biggr)/(v_++v_-)$, and
$v_\pm=(u_{\rho 1}^2+u_{\rho 2}^2)/2\pm
\sqrt{(u_{\rho 1}^2-u_{\rho 2}^2)^2+4b^2}/2$.

The criterion
indicating strong superconducting fluctuations in a given chain $i$
is then translated
from Eqs. (\ref{criterion singlet}) and (\ref{criterion triplet})
using the definitions for $A_{i j}$ and $B_i$.
The strong superconducting fluctuations
induced by the interchain coupling could be observed
in wires where only two channels propagate, or alternatively
in one--dimensional coupled electron--hole systems.
There are two
additional
correlation functions describing superconducting fluctuations
for this particular system.
These occur when a singlet or triplet
Cooper pair is formed with one electron
from chain 1 and one electron from chain 2. They are
ignored here because the
criteria for Cooper pairs formation within one branch
are in general satisfied for a lower threshold phonon coupling.

The possibility of superconducting transitions
in two channel systems could  in principle be probed e.g.
by studying the
persistent current of a two channel Mesoscopic ring.
Rings with few channels built from GaAs/AlGaAs
heterostructures have been recently
fabricated to study the persistent current in the
quasi ballistic regime \cite{Mailly}.
In the normal metal regime, the Fourier transform of the power spectrum
\cite{Webb single flux quantum} yields in general
two peaks: one associated with the flux quantum, signaling the twist
in the boundary condition of the electron wave function due to
the flux which threads the ring, and another, smaller peak
at half the flux quantum associated with weak localization effects.
The flux causes a twist in the boundary conditions, but the
bulk properties (such as the correlation function exponents)
are unaffected. It is plausible that by varying the
electron density with an overall
electrostatic gate, one could induce strong superconducting
fluctuations using the WB mechanism.
In this regime, the peak associated with the single flux quantum
will decrease while the $\phi_0/2$ component will be enhanced due to the
presence of Cooper pairs. Whether or not a strictly two channel ring
can be achieved is still a technological challenge. If disorder associated with
impurities and irregularities in the electrostatic confinement
can be minimized, a two channel ring is effectively achieved just by
``pinching'' the ring with a point contact positioned along the ring.
Moreover, self inductance effects between the two modes
can be safely neglected as these effects have been shown to be proportional to
the fine structure constant multiplied by the ratio of the Fermi velocity
divided by the velocity of light \cite{LossM}.

\subsection{Relevance of hopping between modes}

The relevance of hopping between two {\it chains}
(as opposed to modes) has been discussed on several occasions
\cite{Schulz,Anderson} and there is some controversy regarding this.
While Ref. \cite{Anderson} states that a critical coupling between chains must
be overcome before the Luttinger liquid phase
becomes unstable, Ref. \cite{Schulz} uses a renormalization group argument
to show that transverse hopping is always relevant.

In the situation we are concerned with, however, the hopping bewteen
chains is {\it already} included: the ground state is a
symmetric combination of  the two chain states, and the first excited
state is an antisymmetric combination of these states. As a result,
transitions between  modes will only occur as a result of local
imperfections in the confinement potential, and scattering with impurities.
These processes are thus described by the potential:
\begin{equation}
V_h=\sum_{i, j, s} \sum_k t_k(\psi_{i s}^\dagger(x_k)
\psi_{j s}(x_k)+{\rm H.c.})~,
\end{equation}
with $x_k~(k=1,2,...)$ indicating the position of the imperfection/
impurity along the wire.
The relevance of hopping
between these two modes does not seem to cause a problem. However,
a rigorous treatment should be provided,
and is presently under way.

\section{Transport properties}

We discuss the transport properties of a 1 D
interacting electron system coupled to phonons, first for an ideal wire, and
then in the limit of a weak scatterer and a weak link.
Our analysis follows the work of Ref. \cite{Kane}.

\subsection{Conductivity of an ideal wire}

The conductivity is given by the Kubo formula:
\begin{eqnarray}
\sigma(q,\omega)={1\over i \omega}
\biggl[
&i&\int dx\int dt~ e^{i\omega t-iqx}<j_D(x,t)j_D(0,0)>\nonumber\\
&-&{2\over \pi} u_\rho K_\rho\label{Kubo}
\biggr]~,
\end{eqnarray}
With the second term representing the contribution of the
diamagnetic part of the current operator,
and $j_D\equiv\sqrt{2/\pi}u_\rho K_\rho\Pi_\rho$ the paramagnetic current.
To calculate this quantity, it is useful to
perform a Wick rotation to imaginary times. The calculation follows
that of Ref. \cite{Shankar} for spinless Fermions.
The diamagnetic part is cancelled by contributions coming from the second term
in Eq. (\ref{Kubo}), and the only remaining contribution is
proportional to
$<\partial_\tau\varphi_\rho(x,\tau)\partial_\tau\varphi_\rho(0,0)>$.
At zero temperature, the conductivity becomes:
\begin{equation}
\sigma(q,\omega)={2iu_\rho K_\rho\over\bar{\omega}
}\biggl({\bar{\omega}^2(\bar{\omega}^2+c^2q^2)\over(\bar{\omega}^2+v_+^2q^2)
(\bar{\omega}^2+v_-^2q^2)}\biggr)_{\bar{\omega}=i\omega}
\end{equation}
which can be compared to the uncoupled case (no phonons):
\begin{equation}
\sigma_0(q,\omega)={2iu_\rho K_\rho\over\bar{\omega}
}\biggl({\bar{\omega}^2\over
\bar{\omega}^2+u_\rho^2q^2}\biggr)_{\bar{\omega}=i\omega}
\end{equation}
Note that for large momentum, the conductivity does not depend
on the phonon parameters:
\begin{equation}\lim_{q\rightarrow 0}\sigma(q,\omega)
={2u_\rho K_\rho\over\pi}\biggl[i{\cal P}(\omega)+\pi\delta(\omega)\biggr]
{}~.
\end{equation}
In the DC case, a finite quantity can therefore be obtained
by considering the quantity $\lim_{\omega\rightarrow 0} \omega\sigma(\omega)$.

The result for the zero temperature DC conductivity
are rather dull: at finite temperatures, however, effects
associated with the phonons should survive,
even in the limit $q\rightarrow 0$. The finite temperature dependence
can be obtained from conformal invariance arguments \cite{Affleck},
but will not be considered here.

\subsection{Conductance of an ideal wire}

We now consider a
quantity which can be obtained directly
from an $I-V$ measurement on a small wire: the conductance.
The Landauer formula \cite{Landauer}
provides a simple way to relate the conductance
of a Mesoscopic sample to the quantum mechanical transmission properties
of this sample. While some generalizations of this formula to
specific systems of interacting electrons have been proposed \cite{Meir
Wingreen},
its main application is for non interacting
electrons. Fisher and Lee \cite{Fisher Lee} have introduced a general
expression relating the transmission properties of the sample to the
Greens function: this approach assumes that a constant electric field
is applied to a portion of wire with  a finite length $L$, and calculates
in the linear response regime the resulting current in the DC limit.
This approach has drawn some criticism \cite{Landauer criticism}
because it does not describe an open system (the leads to which the sample
is connected are not taken into account).
Nevertheless, the Fisher Lee formula will be used here,
as it provides a systematic way to generalize single electron
results to arbitrary
many body systems. The end product can be compared to the free electron case
at any stage of the calculation. The Fisher--Lee
formula reads:
\begin{eqnarray}
G=\lim_{\omega\rightarrow 0}{1\over\hbar\omega L}\int d\tau\int_0^L dx
&<&T_\tau\partial_\tau\varphi_\rho(x,\tau)
\partial_\tau\varphi_\rho(0,0)>\nonumber\\
&\times& e^{i\omega\tau}~.\label{Fisher Lee}
\end{eqnarray}
The thermodynamic limit, if necessary, is taken only once the DC limit
$\omega\rightarrow 0$ has been specified.

For a Luttinger liquid coupled to phonons, the result is:
\begin{equation}
G={2e^2\over h}K_\rho A~.\label{conductance}
\end{equation}
Note that since $A>1$, the conductance of this system is always enhanced by the
phonon coupling. In particular, for free electrons with $K_\rho=1$,
the conductance associated with an electron channel is larger than
that associated with the conductance quantum. Using once again
the analogy with coupled electron systems of Sec. \ref{analogy}, we could
envision an experiment
performed in the integer quantum Hall regime, where the number of
edge states propagating on the boundaries of the sample correspond to the
number of filled bulk Landau levels. Edge states can be described
by chiral Luttinger liquids \cite{Wen}, as for a given edge the
direction of propagation is fixed by the $E\wedge B$ drift
(E is the electric field associated with the confining potential).
In the IQHE, each (isolated) edge state
has a Fermi liquid behavior because
the Luttinger liquid parameter $K_\rho$ takes
the marginal value $K_\rho^*=1$. In a situation with only two Landau
level filled, and assuming that the separation between edge states
1 and 2 is comparable to the magnetic length (so that the interaction between
edge states is noticeable),
Eq. (\ref{conductance}) predicts
a conductance which is {\it larger} than the free electron value!

\subsection{Weak Barriers}

The effects of impurity scattering for a weak barrier can be analyzed
following the perturbative renormalization group treatment of Kane and Fisher
\cite{Kane}. Weak scattering is described by an additional term in the
Hamiltonian:
\begin{equation}
\delta H=\sum_s\int dx ~v(x)\psi_s^\dagger(x)\psi_s(x)
\end{equation}
For simplicity, we assume that $v(x)$ is short ranged and centered at $x=0$,
so that the fields
away from the impurity can be integrated out. Moreover, it was
argued in Ref. \cite{Kane} that the renormalization group will generate
higher order terms in the perturbation series so that the most general
effective action associated with the perturbation becomes:
\begin{eqnarray}
\delta S=\int d\tau\sum_{n_\rho+n_\sigma {\rm even}}&&v(n_\rho, n_\sigma)
\label{perturbed action}\\
&&\times\exp(\sqrt{\pi}(n_\rho\varphi_\rho(0)+n_\sigma\varphi_\sigma(0))
\nonumber\end{eqnarray}
where $v(n_\rho, n_\sigma)$ is the coupling strength associated with the
transfer  of $n_\rho$ electron charges and $n_\sigma$ electron
spins accross the weak barrier.
A perturbative renormalization group analysis of the partition function then
leads to the flow equations:
\begin{equation}
{\partial v\over\partial l}(n_\rho, n_\sigma)=\biggl(
1-{n_\rho^2\over 2} AK_\rho
-{n_\sigma^2\over 2}\biggr)v(n_\rho, n_\sigma)~.
\label{flow v}\end{equation}
In particular, in the case where on electron is transfered, we see that
$v_{11}$ is relevant when $AK_\rho<1$, and the system flows towards
the strong coupling (large barrier) behavior. As was seen in Sec.
\ref{2 correlation}, $AK_\rho<1$ corresponds to the criterion for SDW (CDW).
The temperature dependence of the conductivity can be obtained
perturbatively:
\begin{equation}
G(T)={2e^2\over h}\biggl(AK_\rho-\sum_{n_\rho, n_\sigma}c(n_\rho, n_\sigma)
T^{n_\sigma^2+n_\rho^2AK_\rho-2}+...\biggr)
\label{temperature weak}
\end{equation}
This expression seems to imply that the conductivity diverges at low
temperature
in the SDW (CDW) regime. However, this is the regime where the perturbation
expansion breaks down, and we have to use a strong barrier theory
to analyse this case.

\subsection{Strong Barriers}

The strong barrier situation can also be addressed
because of the duality properties
of the action. The action can be written in terms of the field
$\varphi_\nu$, or alternatively in terms of the fields
$\theta_\nu(x,\tau)=\int^ x dx^\prime
\Pi_\nu(x^\prime,\tau)$. A large barrier is described by the hopping term:
\begin{equation}
\delta h=\sum_st(\psi_{+s}^\dagger(x=0)\psi_{-s}+{\rm H.c.})~.
\end{equation}
Including the multiple hopping processes,
the additional term in the action becomes \cite{Kane}:
\begin{eqnarray}
\delta S=\sum_{n_\rho+n_\sigma {\rm even}}t(n_\rho, n_\sigma)\int d\tau
{}~&&\cos(n_\rho\sqrt{\pi}\Pi_\rho)\nonumber\\
&&\times\cos(n_\sigma\sqrt{\pi}\Pi_\sigma)~,
\end{eqnarray}
with the fields evaluated at $x=0$.
The perturbative renormalization group analysis then generates the flow
equations:
\begin{equation}
{\partial t\over \partial l}(n_\rho, n_\sigma)=
\biggl(1-{n_\rho^2\over 2}{B\over K_\rho} -{n_\sigma^2\over 2}\biggr)
t(n_\rho, n_\sigma)
\end{equation}
Again for the case of single electron hopping, $t_{11}$ becomes relevant when
$B/K_\rho<1$, which corresponds to the superconducting
regime of Sec. \ref{2 correlation}.
The temperature dependence of the conductance is given by:
\begin{equation}
G(T)={2e^2\over h}\sum_{n_\rho, n_\sigma} c^\prime(n_\rho, n_\sigma)
T^{(B/K_\rho)n_\rho^2+n_\sigma^2-2}+...
\end{equation}

\subsection{Phase diagram}

It is tempting to unify the weak and strong barrier regimes in the
phase diagram of Fig. \ref{fig4} a). In this figure, the dashed line at
$K_\rho=1$ represents the marginal line which separates the strong and and weak
relevant regime for electrons with no phonons in Ref. \cite{Kane}.
The dashed--dotted lines give the direction of the flow.
The introduction of phonon coupling modifies the phase diagram
significantly.
First of all, the region where strong coupling is relevant shifts towards
lower values of $K_\rho$, as the coupling to phonons compensates partly the
repulsive interaction between electrons. Because $AB>1$, the regions where
$v$ and $t$ are irrelevant overlap, in sharp contrast with the case where no
phonons are present.
This is the manifestation of the intermediate (M) phase, where it was shown for
the Hubbard model that no response function diverges.
We therefore expect three different regimes:
a) the SDW (CDW) regime, where weak scattering is relevant and strong barriers
are irrelevant. The system is an insulator, as the conductance $G(T)\sim
T^{B/K_\rho-1}$ vanishes at low temperature. The I-V characteristic  is then
non linear, $I(V)\sim V^{B/K_\rho}$.
b) the superconducting regime, where weak scattering is irrelevant, and strong
barriers are relevant perturbations: at low temperatures, the conductance is
essentially that of an ideal wire, and the I-V characteristic is linear.
c) the intermediate (M) regime, where it is unfortunately not possible to get
an analytic expression for the temperature dependence of the conductance.
The conductance will behave either as in case a) or as in case b), depending
on the value of $K_\rho$ and the coupling constants $v$ and $t$. For
$1/A<K_\rho<B$, it is at present not possible to obtain the exact
location of the marginal line, because no rigorous treatment
seems possible in this region.

As one tunes the parameters of the system
close to the WB singularity, some
spectacular effects are expected. This case
is illustrated in Fig.
\ref{fig4} b): at low $v$, the marginal line moves all the
way to $K_\rho=0$, while
at low $t$ it stops moving at $K_\rho=B^*$, with $B^*$ defined in Sec.
\ref{Green subsection}. Weak coupling becomes irrelevant regardless
of the strength of the electron electron interaction, and the conductance
diverges.

Note that by considering the transport properties
of the coupled electron-phonon system, we have in fact recovered all the
physics derived with the help of the correlation functions of
Sec. \ref{2 correlation}. The transport properties thus provide
us with a diagnostic tool to analyse the ``phase'' specified
by the parameters of the model, as the perturbative renormalization group
yields exact results in this particular case.

\section{Conclusion}

We have analyzed the role of the WB singularity in a system
of interacting electrons coupled to phonons in 1D. The Luttinger
liquid description allowed us to get exact results for the correlation
functions which in turn determine the dominant type of fluctuations. For the
Hubbard model, the competition between the repulsive interaction and the
attractive
retarded interaction mediated by the phonons has been clearly displayed.
An intermediate phase separates the SDW (CDW) from the superconducting
regimes. By increasing the electron density towards half filling, the presence
of the WB singularity suppresses SDW (CDW) order and pushes
the system towards the superconducting phase. This could be
checked in systems where two electron branches are coupled
to each other, where complications associated
with $2k_F$ retarded phonon processes are absent.
The coupling to phonons affects the transport properties, enhancing the
conductivity, and suggesting that it is possible to achieve in certain
cases a conductance
per channel larger than the conductance quantum. The analysis of the
effect of disorder allows to recover the three regimes predicted by
the correlation functions: an insulating regime and a superconducting regime
(with finite conductance) separated by an intermediate phase.
The insulating regime disappears totally at the WB point, and the conductance
becomes infinite.

There are still many standing issues. First of all, is it conceivable
that the WB singularity survives an additional $2k_F$ phonon
interaction? For electron  bandwidths which are sufficiently small that the
retardation effects of $2k_F$ phonons can be neglected, the answer is
yes, because then the on site repulsion parameter $U$
is simply shifted towards lower values. Nevertheless, in the general
case, no answer is available at the moment.
Second, it seems quite plausible that the WB singularity could
exist also in higher dimensional systems. However, in general
2 and 3 D systems do not possess the luxury of an exact solution, which
renders the analysis more difficult. Finally, the analogy with
electron wave guides suggests that it would also be interesting to study
the WB singularity for a wire with an arbitrary number of coupled modes.

\begin{figure}
\caption{Green function exponent for the Hubbard model:
a) at fixed filling factor, as a function of $u=U/t$ for
$b/cu_\rho=0$ (solid line), $b/cu_\rho=0.2$ (dotted line),
$b/cu_\rho=0.4$ (dashed line) and $b/cu_\rho=0.6$ (dashed--dotted line).
b) for fixed $u$, as a
function of filling factor and for the same values of $b/cu_\rho$ as in
a).}\label{fig1}
\end{figure}
\begin{figure}
\caption{a) Phase diagram of the Hubbard model coupled to phonons,
as a function of filling factor and phonon coupling constant $b/cu_\rho$,
for $U/t=0.3$.
b) Phase diagram of the Hubbard model coupled to phonons
at quarter filling ($n=1/2$), as a function of $U/t$ and $b/cu_\rho$.}
\label{fig2}
\end{figure}
\begin{figure}
\caption{Plot of $u_\rho/K_\rho$ as a function of the filling factor $n$.
On the left hand side, we have from top to bottom $U=16,8,4,2$.
Note the abrupt change for small values of $U$ as one approaches half
filling.}\label{fig3}
\end{figure}
\begin{figure}
\caption{Flow diagram for the transport properties of a Luttinger
liquid coupled to phonons: a) for an arbitrary electron--phonon coupling.
$K(=K_\rho)$ is the Luttinger liquid parameter, $v$ is the weak barrier
coupling, and $t$ is the hopping strength. b) same as a), but for
parameters corresponding to the WB singularity
($A\rightarrow\infty$ here).}\label{fig4}
\end{figure}

\begin{references}
\bibitem[*] {} Permanent address as of October 1st, 1994: Institut Laue
Langevin, B.P. 156, 38042 Grenoble, France.
\bibitem{Rice} W. Apel and T. M. Rice, Phys. Rev. B {\bf 26}, 7063 (1982).
\bibitem{Altschuler} B.L. Altschuler, A.G. Aronov and D.E. Khmelnitsky,
J. Phys. C {\bf 15}, 7367 (1982).
\bibitem{Ambegaokar} V. Ambegaokar and U. Eckern, Phys. Rev. Lett.
{\bf 65}, 381 (1990).
\bibitem{Coulomb} see for example: C.W.J. Beenakker and H. van Houten,
in {\it Single charge tunneling}, H. Grabert and M.H. Devoret eds.
(Plenum, New York 1991).
\bibitem{Emery} V. J. Emery, in {\it Highly Conducting One
Dimensional Solids}, J. T. Devreese, R. P. Evrard and V. E. van Doren
eds.  (Plenum, New York, 1979), p.327.
\bibitem{Solyom} J. S\'olyom, Adv. Phys. 28, 201 (1979).
\bibitem{Mattis} D. C. Mattis and E. H. Lieb, J. Math. Phys. {\bf 6},
304  (1965).
\bibitem{Luther} A. Luther and I. Peschel, Phys. Rev. B {\bf 9},
2911  (1974).
\bibitem{Haldane}
F. D. M. Haldane, J. Phys. C {\bf 14}, 2585  (1981);
Phys. Rev. Lett. {\bf 47}, 1840  (1981).
\bibitem{Kane} C. L. Kane and M. P. A. Fisher, Phys. Rev. Lett.
{\bf 68}, 1220  (1992); Phys. Rev. B. {\bf 46}, 7268 (1992); Phys. Rev. B
{\bf 46}, 15233 (1992).
\bibitem{Furusaki} A. Furusaki and N. Nagaosa, Phys. Rev. B {\bf 47},
4631 (1993); Phys. Rev. B {\bf 47}, 3827 (1993).
\bibitem{Moon} K. Moon, H. Hi, C.L. Kane, S.M. Girvin and M.P.A. Fisher,
Phys. Rev. Lett. {\bf 71}, 4381 (1993).
\bibitem{Webb} F.P. Milliken, C.P. Umbach and R.A. Webb, preprint (1994).
\bibitem{Kane noise} C.L. Kane and M.P.A Fisher, Phys. Rev. Lett. {\bf 72},
724 (1994).
\bibitem{Kondo} D.H. Lee and J. Toner, Phys. Rev. Lett. {\bf 69}, 3378
(1992); A. Furusaki and N. Nagaosa, Phys. Rev. Lett. {\bf 72}, 892 (1994).
\bibitem{Loss} D. Loss, Phys. Rev. Lett. {\bf 69}, 4630 (1992).
\bibitem{LossM} D. Loss and Th. Martin, Phys. Rev. B. {\bf 47},
4619 (1993).
\bibitem{Wentzel} G. Wentzel, Phys. Rev. {\bf 83}, 168 (1951).
\bibitem{Bardeen} J. Bardeen, Rev. Mod. Phys. {\bf 23}, 261 (1951).
\bibitem{Engelsberg} S. Engelsberg and B. B. Varga Phys.Rev. {\bf 136},
A1582 (1964).
\bibitem{Voit0} J. Voit and H. J. Schulz, Mol. Cryst. Liq. Cryst.
{\bf 119}, 449 (1985).
\bibitem{Apostol} M. Apostol and I. Baldea, Phys. Lett. {\bf 88A},
73 (1982); J. Phys. C {\bf 15}, 3319 (1982).
\bibitem{Voit1} J. Voit and H. J. Schulz, Phys. Rev. B {\bf 34}, 7429
(1986); {\bf 36}, 968 (1987); {\bf 37}, 10068 (1988).
\bibitem{Zimanyi} G. T. Zimanyi, S. A. Kivelson and A. Luther,
Phys. Rev. Lett. {\bf 60}, 2089 (1988).
\bibitem{Voit2} J. Voit, Phys. Rev. Lett. {\bf 64}, 323 (1990).
\bibitem{Fetter} A. Fetter and D. Walecka, {\it Many Particle
Physics} (McGraw--Hill, New York, 1969).
\bibitem{Schulz} H. J. Schulz, Phys. Rev. Lett. {\bf 64}, 2831 (1990);
Int. J. of Mod. Phys.   B {\bf 5},
57  (1991).
\bibitem{Loss Martin} D. Loss and Th. Martin, to appear in Phys. Rev. B
(1994).
\bibitem{Martin Loss} Th. Martin and D. Loss, to appear in Int. J. Mod.
Phys. (1994).
\bibitem{footnote1} This singularity was left unnoticed
in Ref. \cite{Voit0} (only forward scattering is included there),
and the exponents given there seem to be incorrect.
\bibitem{Logarithm} T. Giamarchi and H.J. Schulz, Phys. Rev. B
{\bf 39}, 4620 (1989).
\bibitem{Lieb} E. H. Lieb and F. Y. Wu, Phys. Rev. Lett. {\bf 20}, 1445
(1968).
\bibitem{Coll}C. F. Coll, Phys. Rev. B {\bf 9},2150 (1974).
\bibitem{footnote} Note that the model becomes unphysical
when we go  below the WB singularity
$u_\rho^*/K_\rho^*$ in Fig. 2 by increasing
the filling factor towards half filling.
\bibitem{Van Wees} B. Van Wees {\it et al.}, Phys. Rev. Lett. {\bf 62},
1181 (1988).
\bibitem{Matveev} K. A. Matveev and L. I. Glazman, Phys. Rev. Lett.
{\bf 70}, 990 (1993).
\bibitem{Penc} K. Penc and J. S\'olyom, Phys. Rev. B {\bf 47},
6273 (1993).
\bibitem{Mailly} D. Mailly, C. Chapelier, and A. Benoit,
Phys. Rev. Lett. {\bf 70}, 2020 (1993).
\bibitem{Webb single flux quantum} R.A. Webb, S. Washburn, C.P. Umbach
and R.B. Laibowitz, Phys. Rev. Lett. {\bf 54}, 2696 (1985).
\bibitem{Anderson}P. W. Anderson, Phys. Rev. Lett. {\bf 67}, 3844
(1991).
\bibitem{Shankar} R. Shankar, Int. J. Mod. Phys. B {\bf 4}, 2371 (1990).
\bibitem{Affleck} S. Eggert, I. Affleck and M. Takahashi, preprint (1994).
\bibitem{Landauer} R. Landauer, Phil. Mag. {\bf 21}, 863 (1970).
\bibitem{Meir Wingreen} Y. Meir and N.S. Wingreen, Phys. Rev. Lett.
{\bf 68}, 2512 (1992).
\bibitem{Fisher Lee} D.S. Fisher and P.A. Lee, Phys. Rev. B
{\bf 23}, 6851 (1981).
\bibitem{Landauer criticism} R. Landauer, Phys. Scripta {\bf T242},
110 (1992).
\bibitem{Wen} X.G. Wen, Phys. Rev. B {\bf 43}, 11025 (1991); Phys. Rev. Lett.
{\bf 64}, 2206 (1990); Phys. Rev. B {\bf 44}, 5708 (1991).
\end{references}
\end{document}